\documentclass[aps,prl,twocolumn,superscriptaddress,showpacs,floatfix]{revtex4-2}
\usepackage{graphicx,bm,epsfig,textcomp,color,dcolumn,setspace,array,mathrsfs,amsmath,amssymb,gensymb,booktabs,url,hyperref,bibentry,natbib}
\begin{document}

\title{Quantum interference in a superconductor-${\mathrm{MnBi}}_{2}{\mathrm{Te}}_{4}$-superconductor Josephson junction}

\author{Yu-Hang Li}
\thanks{yuhang.li@ucr.edu}
\affiliation{Department of Electrical and Computer Engineering, University of California, Riverside, California 92521, USA}
\author{Ran Cheng}
\thanks{rancheng@ucr.edu}
\affiliation{Department of Electrical and Computer Engineering, University of California, Riverside, California 92521, USA}
\affiliation{Department of Physics and Astronomy, University of California, Riverside, California 92521, USA}

\begin{abstract}
We study the transport properties of a Josephson junction consisting of two identical $s$-wave superconductors separated by an even-layer ${\mathrm{MnBi}}_{2}{\mathrm{Te}}_{4}$ (MBT). Using recursive Green's function method, we calculate the supercurrent in the presence of a perpendicular magnetic field and find that its quantum interference exhibits distinct patterns when the MBT is in different magnetic states. In the antiferromagnetic state, the MBT is an axion insulator supporting an extended ``hinge" supercurrent, which leads to a sinusoidal interference pattern decaying with the field strength. In the ferromagnetic state, the MBT is a Chern insulator and the unbalanced chiral supercurrents on opposite edges give rise to a highly asymmetric interference pattern. If the MBT turns into a metal as the Fermi level is tuned into the conduction band, the interference exhibits a Fraunhofer pattern due to the uniformly distributed bulk supercurrent. Our work unravels a strong indicator to identify different phases in the MBT and can be verified directly by experiments.
\end{abstract}

\maketitle

The inquiry into exotic topological phases protected by symmetries has never stagnated in the past decades~\cite{Hasan2010Coll,Qi2010Topo,Ando2013Topo}. Recently, an emerging direction dubbed intrinsic magnetic topological insulators in which magnetism and topology of electrons are intertwined has witnessed a surge of interest~\cite{Tokura2019Magnetic,Sekine2021Axion,Zhao2021Routes,Nenno2020Axion}. For example, an even layer ${\mathrm{MnBi}}_{2}{\mathrm{Te}}_{4}$ (MBT) thin film exhibits strong correlation between its electronic phases and magnetic configurations~\cite{Li2019Intrinsic,Gu2021Spectral,Zhang2019Topo}. In the antiferromagnetic ground state, the MBT is believed to be an axion insulator (AI) with a quantized axion field $\theta=\pi$ protected by the combined parity and time reversal ($\mathcal{PT}$) symmetry~\cite{Qi2008Topological,Qi2010Topo}. In the ferromagnetic state, it becomes a Chern insulator (CI) characterized by a quantized Chern number with its sign determined by the magnetization direction~\cite{Liu2020Robust}.

The CI phase in the MBT has been widely confirmed by experiments~\cite{Deng2020Quantum,Cai2022Electric}, where the transport measurement finds a quantized Hall conductance along with a vanishing longitudinal one on a Hallbar structure. However, although some transport experiments have hinted at its existence~\cite{Liu2020Robust}, the AI phase in the ground state remains elusive since the observed large resistance and vanishing Hall effect are common behaviors of normal insulators. In addition to the striking topological magnetoelectric effect~\cite{Qi2008Topological,Li2022Iden}, the half-quantized layer Hall effect ascribing to the surface Chern number is also deemed a fingerprint of an AI~\cite{Nomura2011Surface}. Nevertheless, the surface Hall conductance reported in a recent experiment deviates significantly from the anticipated value of $e^2/(2h)$~\cite{Gao2021Layer}. Contrary to that in a CI, the nonlocal transport in the AI phase is found to be diffusive rather than ballistic~\cite{Li2021Mate}, as the edge current is substantially extended into the bulk~\cite{Lin2022Infl}. These unexpected behaviors imply the existence of ``hinge" states in an AI~\cite{Gu2021Spectral,Gong2022Half}.

In this paper, we resolve this puzzle by studying the transport properties of a Josephson junction involving an even layer MBT as the tunnel barrier. We find that the supercurrent in the AI phase is symmetric along opposite edges and intrinsically much wider than that in the CI phase thanks to the backscattering between counterpropagating ``hinge" states on each edge. In the presence of a perpendicular magnetic field, the distinct distribution of supercurrent results in different quantum interference patterns that can serve as a strong indicator of different magnetic states; hence, different electronic phases in the MBT. In the AI phase, the quantum interference exhibits a decreasing harmonic pattern symmetric about the field direction. In the CI phase, the interference displays a highly asymmetric harmonic pattern attributed to the unbalanced supercurrent propagating oppositely on different edges. As a comparison, when the Fermi energy lies inside the conduction band (i.e., normal metal phase), the supercurrent is uniformly distributed inside the bulk, leading to a typical Fraunhofer pattern in the quantum interference.

We start with the low energy effective Hamiltonian for an even-layered MBT~\cite{Zhang2019Topo,Li2022Iden}
\begin{align}
H_{\rm MBT}=\sum_{a=0}^3d_a(\bm{k})\Gamma_a+\sum_{z=1}^{L_z}M(z)\sigma_3\otimes \tau_0,
\label{MBT_Ham}
\end{align}
where the first term describes a 3D topological insulator with $d_0(\bm{k})=M_0-B_1k_z^2-B_2(k_x^2+k_y^2)$, $d_{1(2)}(\bm{k})=A_2k_{x(y)}$ and $d_3(\bm{k})=A_1k_z$, and the second term describes the exchange interaction between the topological electrons and the magnetizatic moments in each layer. Here, we have adopted the basis $\psi_{\bm{k}}=(c_{s\bm{k}\uparrow},\ c_{p\bm{k}\uparrow},\ c_{s\bm{k}\downarrow},\ c_{p\bm{k}\downarrow})^T$ with $c_{\lambda\bm{k}\sigma}$ the annihilation operator, under which the Dirac $\Gamma$ matrices $\Gamma_{(0,1,2,3)}=(\sigma_0\otimes\tau_3, \sigma_1\otimes\tau_1,\sigma_2\otimes\tau_1, \sigma_3\otimes\tau_1)$. The crystal momentum $\bm{k}$ is confined in the first Brillouin zone of a $L_x\times L_y\times L_z$ cubic lattice with lattice constant $a_0$. The exchange interaction opens a nontrivial gap on the surface Dirac cone, thus giving rise to a $\pm\frac{1}{2}$ surface Chern number when the Fermi energy lies inside the gap~\cite{Qi2008Topological}. In the antiferromagnetic state, the system is an AI with opposite surface Chern numbers~\cite{Essin2009Magnetoelectric}, In the ferromagnetic state, the system is a CI with the surface Chern numbers adding up to one~\cite{Deng2020Quantum,Liu2020Robust,Cai2022Electric}.  

\begin{figure}[ttt]
  \centering
  \includegraphics[width=0.9\linewidth]{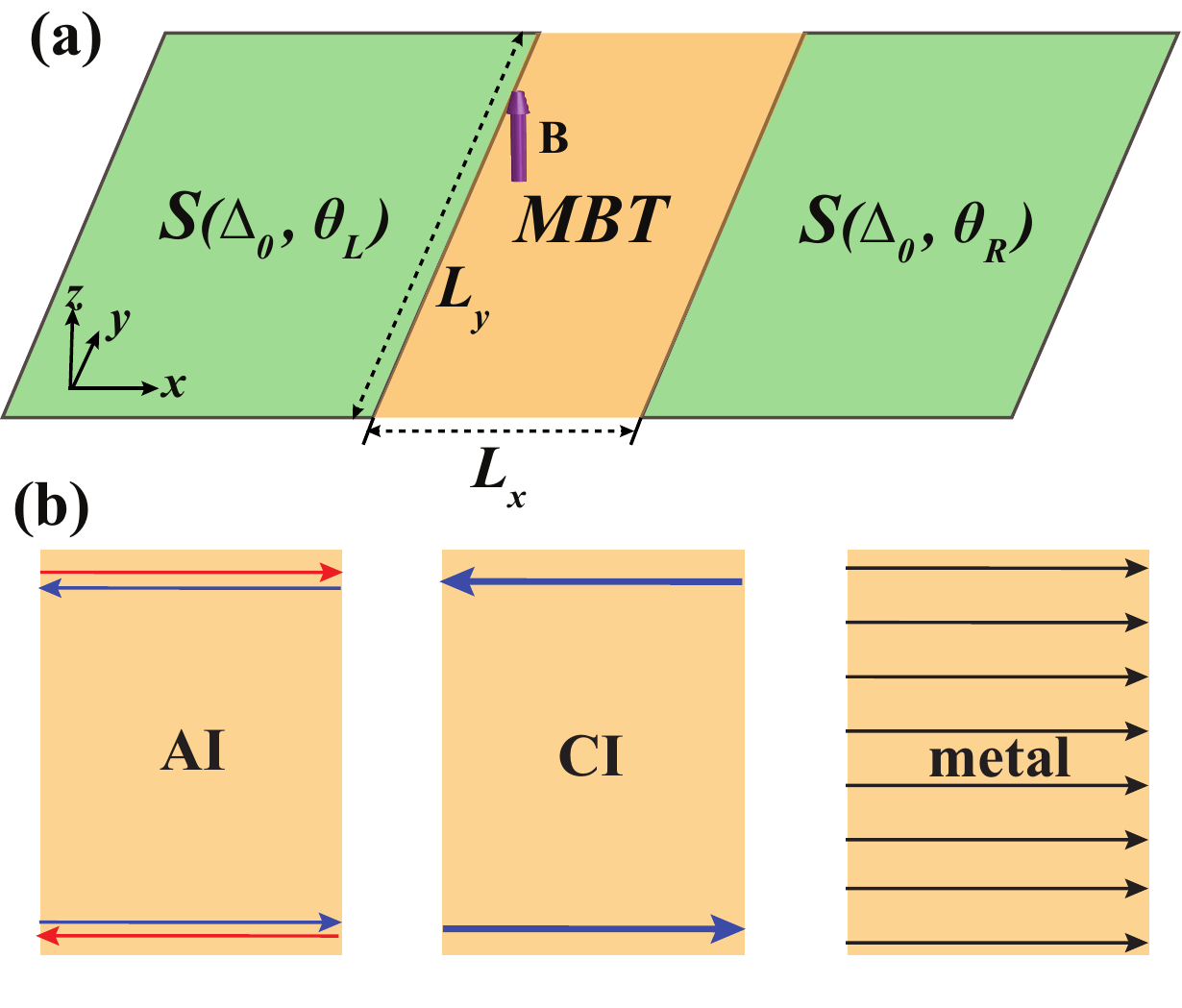}
  \caption{
	(a) Schematic of the MBT Josephson junction. Two $s$-wave superconductors with the same superconducting gap $\Delta_0$ but different phases $\theta_L$ and $\theta_R$ are connected by an even-layer MBT of size $L_x\times L_y$. An external magnetic field $B$ is applied along the $z$ direction. 
	(b) Illustration of the supercurrent distributions in the central MBT region for the AI, CI and metal phases, respectively.
	  	}
\label{sche}
\end{figure}

When sandwiched between two $s$-wave superconductors, the ensuing Bogoliubov-de Gennes Hamiltonian written in the Nambu basis $\Psi_{\bm{k}}=\begin{pmatrix}\psi_{\bm{k}},&\psi_{-\bm{k}}^\dagger\end{pmatrix}$ is
\begin{align}
H_{\rm BdG}=\begin{pmatrix}H_{\rm MBT}(\bm{k})-\mu_M,&&0\\ 0,&&-H_{\rm MBT}^*(-\bm{k})+\mu_M\end{pmatrix},
\label{MBT_BdG}
\end{align}
and the Hamiltonian for each superconductor is 
\begin{align}
H_{\rm SC}=\begin{pmatrix}H_{\rm TI}(\bm{k})-\mu_S,&&\Delta_{\bm{k}}\\ \Delta_{\bm{k}}^{\dagger},&&-H_{\rm TI}^*(-\bm{k})+\mu_S\end{pmatrix},
\label{SC_BdG}
\end{align}
where $H_{\rm TI}(\bm{k})=\sum_{a=0}^3d_a(\bm{k})\Gamma_a$ is the non-magnetic part of Eq.~\ref{MBT_Ham}, $\Delta_{\bm{k}}=i\Delta\sigma_2\otimes\sigma_0$ is the superconducting gap, and $\mu_M$ and $\mu_S$ are the Fermi energies of the MBT and the superconducting leads, respectively. As illustrated in Fig.~\ref{sche}(a), we assume that in the left (right) superconductor, $\Delta_{L(R)}=\Delta_0 e^{i\theta_{L(R)}}$ with $\Delta_0=1$meV and $\mu_S^L=\mu_S^R=200$meV. The exchange gap is $M(z)=\pm50$meV depending on the magnetization direction, and all other parameters in $H_{MBT}$ are taken from Refs.~\cite{Li2022Iden,Zhang2019Topo,Yang2021Odd}. The AI (CI) phase is obtained by setting the magnetization in neighboring layers to be antiparallel (parallel) when the Fermi energy lies inside the exchange gap. The metallic phase, as a control group, can be easily realized by taking $\mu_M=200$meV in the antiferromagnetic state. We stress that an even layer MBT turns into a CI only when the external magnetic field exceeds the spin flop threshold ($\approx3$Tesla)~\cite{Deng2020Quantum,Cai2022Electric,Liu2020Robust}. As a result, robust superconductor with high critical field values such as $\mathrm{MoRe}$ is required to avoid any damage to the superconductivity~\cite{high_Bc}.

The supercurrent across the Josephson junction in Fig.~\ref{sche}(a) is attributed to the Andreev reflection at the interfaces between the MBT and the superconductors. As illustrated in Fig.~\ref{sche}(b), in the AI phase, there are two ``helical'' edge states similar to a quantum spin Hall insulator~\cite{Gu2021Spectral,Gong2022Half}; each edge alone can carry supercurrent. In the CI phase, both edges must be involved because one edge only conducts charges in one direction. In both cases, the supercurrents are in principle carried by the edge states with a finite width. By contrast, when the MBT becomes a metal (for $\mu_M=200$), the supercurrent should be uniformly distributed inside the bulk. The apparent difference in the distribution of supercurrents along the $y$ direction can cause different quantum interference patterns when a perpendicular magnetic field $B$ is applied, which can roughly be anticipated from the Dynes and Fulton method~\cite{Dynes1971Super}
\begin{equation}
I_s(B)=\left|\int_{-L_y/2}^{L_y/2}dyJ_x(y)exp(iyL_xB/2\pi)\right|. \label{eq:DynesFulton}
\end{equation}
where $I_s(B)$ is the total supercurrent, $J_x(y)$ is the supercurrent density flowing in the $x$ direction at vertical location $y$, and $L_x$ and $L_y$ are the system size along $x$ and $y$ directions. We emphasize that Eq.~\eqref{eq:DynesFulton} is oversimplified and cannot deal with supercurrent with $y$ and $z$ components, or uniformity in the thickness direction~\cite{Hui2014Pro}. So we only use Eq.~\eqref{eq:DynesFulton} as a guidance but will not base any real calculations on it.

To calculate the spatial distribution of supercurrent in the MBT, we resort to a rigorous approach---recursive Green's function method. We first discretize the continuum Hamiltonian Eqs.~\eqref{MBT_BdG} and ~\eqref{SC_BdG} on a cubic lattice (lattice constant $a_0$) with open boundary condition. Then the supercurrent flowing from site $\bm{i}$ to $\bm{j}$ inside the central region (the MBT) in Fig.~\ref{sche}(a) can be expressed as~\cite{Akira1994DC,Asano2001Num,Asano2003Jose,Chen2018Asy}
\begin{align}
    I_{\bm{i}\rightarrow\bm{j}}=-\frac{ieT}{\hbar}\sum_n\text{Tr}\{\sigma_3[\hat{T}_{\bm{ij}}\hat{G}_{\omega_n}(\bm{j},\bm{i})-(\bm{i}\leftrightarrow\bm{j})]\},
    \label{RG_super_current}
\end{align}
where $T$($=\Delta_0/20$) is temperature (Boltzmann constant $k_B=1$), $\hat{T}_{\bm{ij}}(=\hat{T}_{\bm{ji}}^\dagger)$ is the hoping matrix between site $\bm{i}$ and $\bm{j}$, and 
the Green's function $\hat{G}_{\omega_n}(\bm{i},\bm{j})=[(i\omega_n-H_{BdG}-\Sigma_L-\Sigma_R)^{-1}]_{\bm{ij}}$ with $\omega_n=(2n+1)\pi T$ the Matsubara frequency. Here, $\Sigma_{L(R)}$ is the self energy due to the coupling to the left (right) superconducting lead, which can be obtained numerically~\cite{Ando1991Qua}. In the presence of an external magnetic field $\bm{B}=B\hat{z}$ inside the central region, the MBT Hamiltonian acquires a Peierls phase $\phi_{ij}=2\pi\int_{\bm{i}}^{\bm{j}}\bm{A}\cdot d\bm{r}/\phi_0$ with $\phi_0=h/(2e)$ being the magnetic flux quanta. To simplify our calculation, we adopt the Landau gauge $\bm{A}=\begin{pmatrix}-yB,&0,&0\end{pmatrix}$. Since the current is conserved inside the central region, the total current can therefore be obtained by summing the current density over the two transverse directions, i.e., $I_s=\sum_{i_y,i_z}I_x({\bm{i}})$. 

\begin{figure}[ttt]
  \centering
  \includegraphics[width=\linewidth]{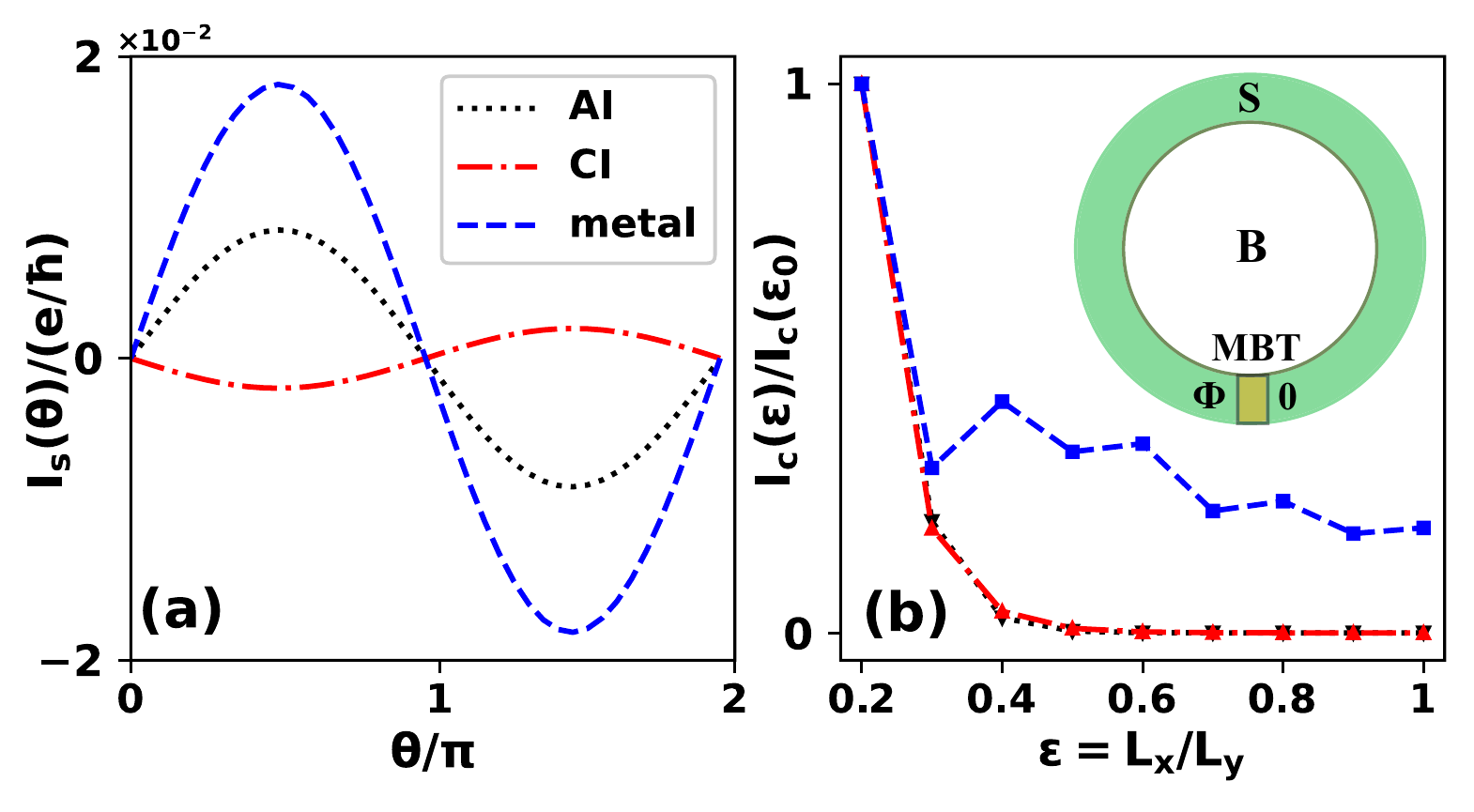}
  \caption{
	(a) Current-phase relations for the MBT Josephson junction in the AI (black dashed line), the CI (red dashed line) and the metal (blue dashed line) phases. System size: $L_x=41$, $L_y=100$, $L_z=6$. The value if $I_s$ in the metal phase is minimized by 20 times for visual clarity. 	
	(b) Critical Josephson supercurrent $I_c$ as a function of the aspect ratio $\varepsilon$ (plotted in the same color scheme). The dashed lines are guides to the eye. Inset: schematic of a superconducting quantum interference device with $L_x=41$, $L_y=100$ and $L_z$=6. 
	  	}
\label{cpr_ld}
\end{figure}

Using Eq.~\ref{RG_super_current} in the absence of magnetic field $B$, we first obtain the current-phase relation, $I_s=I_s(\theta)$ with $\theta=\theta_L-\theta_R$ the quantum phase difference between the two superconducting leads, in a superconducting quantum interference device [SQUID as shown in the inset in Fig.~\ref{cpr_ld}(b)]. For a system size of $L_x\times L_y\times L_z=41\times100\times6$, the results are shown in Fig.~\ref{cpr_ld}(a) for three distinct phases of the MBT. All three cases exhibit a perfect sinusoidal current-phase relation with a $2\pi$ periodicity, i.e., $I_s(\theta)=I_c\sin{(\theta+\theta_0)}$ where $I_c$ is the critical supercurrent and $\theta_0$ is the initial phase, confirming the short Josephson junction limit in all phases that is valid for $L_x\ll\xi$ with $\xi$ the coherence length in the superconductor leads~\cite{Lee2015Ult}. It can roughly be estimated by the equation $\xi=\hbar v_f/\pi \Delta_0$ with $v_f$ the Fermi velocity,  which is about $\xi\sim 1000a_0$ in our model. We notice that the sinusoidal current-phase relation persists even when $L_x=L_y$ for all three cases. Besides, we see that the system is a $\pi$ Josephson junction ($\theta_0=\pi$) when the MBT is a CI while it becomes a $0$ Josephson junction ($\theta_0=0$) in other two phases. Consequently, the MBT enables a platform for a switchable $0$--$\pi$ Josephson junction controlled by a magnetic field~\cite{Gingrich2016}, which holds potential in quantum computing applications~\cite{herr2012josephson,holmes2013energy}. Note that the phase shift $\theta_0$ in the CI phase can also be tuned by the in plane Zeeman energy or random impurities~\cite{Sakurai2017Tunable}.

We then calculate the critical supercurrent $I_c$ as a function of the aspect ratio $\epsilon=L_x/L_y$. The results, as shown in Fig.~\ref{cpr_ld}(b), are normalized to $I_c(\epsilon_0=0.2)$ at fixed $L_y$($=100$), where the dashed lines are guides to the eye. In spite of the oscillation in the metallic phase, $I_c$ in all three phases decrease exponentially with $\epsilon$, which can remarkably well be fitted by the relation $I_c(\epsilon)=I_c(\epsilon_0)\exp{[-(\epsilon-\epsilon_0)/L_0]}$ where $L_0$ is the decay length. Moreover, the AI and CI phases share almost the same decay length, because the supercurrents in both cases are carried by the ballistic edge states. The nonlocal diffusive transport in the AI phase reported by a recent experiment, on the other hand, may originate from the disorder effect~\cite{Liu2021Diss}.

\begin{figure}[ttt]
  \centering
  \includegraphics[width=\linewidth]{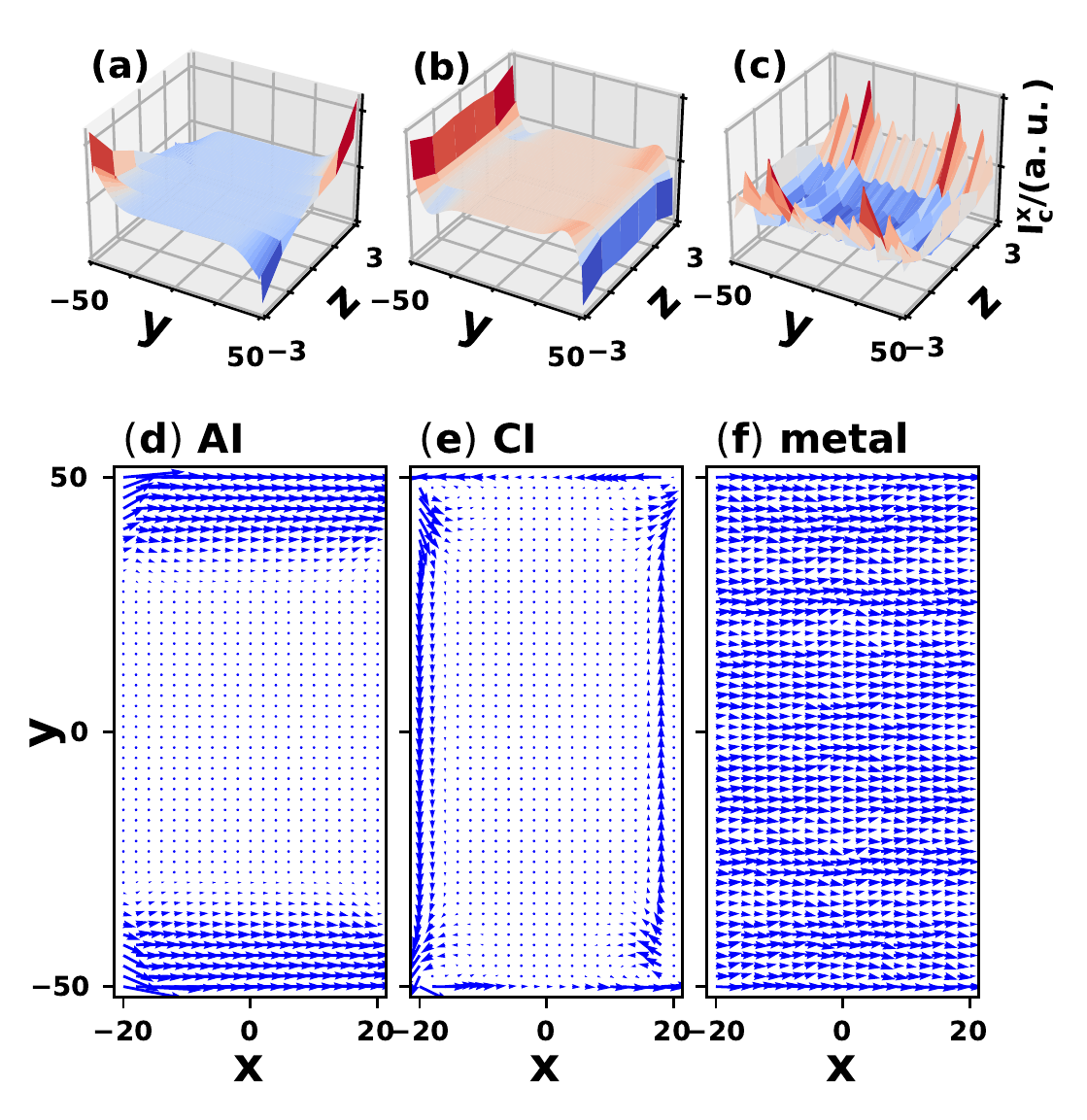}
  \caption{Critical supercurrent distributions for three different phases of the MBT at zero field.
  	(a)-(c) The distribution of the $x$-component on the $y$--$z$ plane $I_x({L_x/2})$ for the AI, CI and metal phases.
  	(d)-(f) In-plane current distributions (vectorial plot) on the $x$--$y$ plane $\bm{I}_{c}(\bm{i})$ for the three corresponding phases.
	System size: $L_x=41$, $L_y=100$, $L_z=6$ and the phase difference is $\theta=\pi/2$ at $B=0$.
	  	}
\label{cu_dis}
\end{figure}

To investigate the quantum interference in the presence of a perpendicular magnetic field, it is instructive to look into the spatial distribution of the critical supercurrent $I_c$ inside the central MBT. Figure~\ref{cu_dis}(a)-(c) plot the distribution of $I_{c}^{x}$ on the cross section of the MBT barrier (i.e., the $y$--$z$ plane) at $x=0$ (center point) for the three different phases. While the superconduct permeates into the bulk in the metallic phase [Fig.~\ref{cu_dis}(c)], it is localized on the two transverse edges with opposite signs in the CI phase [Fig.~\ref{cu_dis}(b)]. In the AI phase, the supercurrent is localized on the boundaries in both $y$ and $z$ directions [Fig.~\ref{cu_dis}(a)], indicating that it is carried by the counter-propagating ``hinge" states on each edge. However, the ``hinge" states in such a thin MBT film cannot be simply regarded as the edge states on the top and bottom surfaces, as they partially overlap in the thickness direction through the conducting side surfaces, leading to substantial back scattering of electrons. Furthermore, the $\mathcal{PT}$ symmetry requires that a pair of diagonal supercurrents [e.g., at $(L_y/2, L_z/2)$ and $(-L_y/2, -L_z/2)$] must have the same magnitude and opposite spin polarizations. This pattern is similar to the helical edge states in a quantum spin Hall insulator~\cite{Liu2008Quantum}. As a result, even one side surface alone (e.g., the one at $L_y/2$) is able to conduct the supercurrent because the top and bottom hinges are counter-propagating. In contrast, both edges (at $\pm L_y/2$) have to be involved in the CI phase in order to form counter-propagating supercurrents.
 
Since the quantum interference depends only on the in-plane supercurrents, we now sum the (critical) supercurrent over all layers and plot the local supercurrent density as an in-plane vector
\begin{align}
 \bm{I}_{c}(i_x,i_y)=\sum_{i_z=-L_z/2}^{L_z/2}\bm{I}_c(\bm{i})
 \label{eq:2dcurrent}
\end{align}
for the three phases of MBT in Figs.~\ref{cu_dis}(d)-(e), one to one corresponding to the upper three panels. Notably, even though the supercurrents in both the AI and CI phases are well localized along the transverse edges, the current distribution in the AI phase is highly extended into the bulk. This wider edge supercurrent can be attributed to the back scattering between the two ``hinge" states at $\pm L_z/2$ as shown in Fig.~\ref{cu_dis}(a). The scattering of hinge currents, hence the broadening of edge currents, has been corroborated by a recent experiment~\cite{Lin2022Infl}. Moreover, the $\mathcal{PT}$ symmetry ensures that the edge supercurrents, after summing over the thickness [see Eq.~\eqref{eq:2dcurrent}], propagate along the same direction and distributes symmetrically along $y$ direction [Fig.~\ref{cu_dis}(d)]. Different from that in the AI phase, the supercurrent in the CI phase propagates along opposite directions on opposite edges. Their magnitudes are the same if the phase difference across the Josephson junction vanishes ($\theta=0$). A nonzero phase difference breaks the balance between the two edge supercurrents, leading to a net flow in the $x$ direction. For the metal phase, we see from Fig.~\ref{cu_dis}(f) that the supercurrent is flat and uniform in the $x$--$y$ plane (with fluctuations). These distinct distributions of in-plane supercurrents will lead to distinct quantum interference patterns.

\begin{figure}[ttt]
  \centering
  \includegraphics[width=0.9\linewidth]{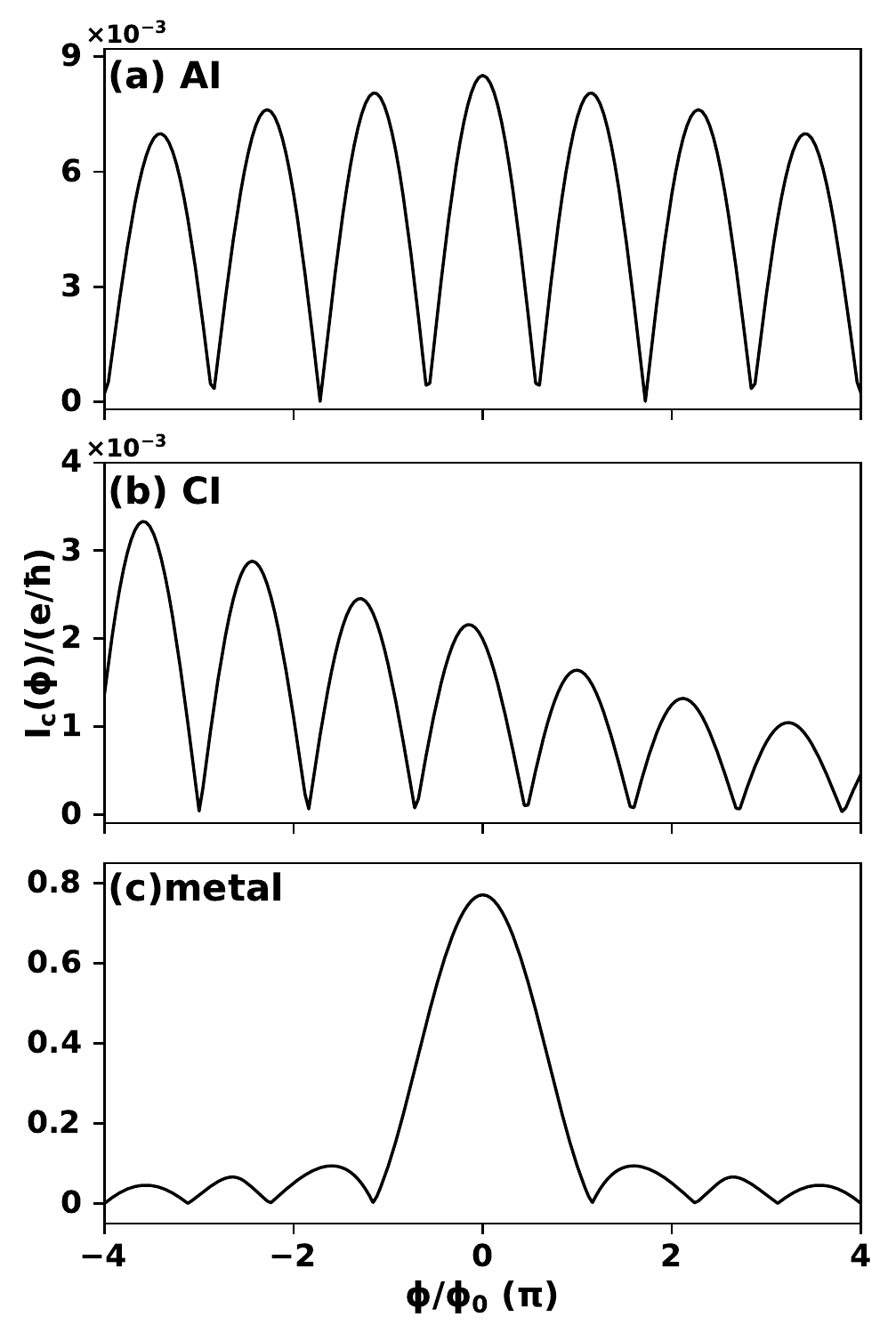}
  \caption{
  	Quantum interference patterns (Critical supercurrent as a function of the magnetic flux) for the MBT in the AI [(a)] CI [(b)] and metal [(c)] phases. Here, the system size is $L_x\times L_y\times L_z=41\times 100\times 6$. 
	  	}
\label{inter}
\end{figure}

Finally, we study the critical supercurrent $I_c$ as a function of the total magnetic flux penetrating through the entire central region $\phi=BL_xL_y$ normalized to the flux quanta $\phi_0=e/(2h)$. Figure~\ref{inter}(a)-(c) plot $I_c(\phi)$ for the AI, CI, and metal phases, respectively, where we indeed observe distinct interference patterns in different phases. In the AI phase, the sinusoidal pattern comes with a period slightly larger than $\phi_0$ and the peaks in each period decrease symmetrically with an increasing magnetic field in both directions, which can be attributed to the symmetrically extended edge supercurrents (consisting of counter-propagating hinge currents) discussed above~\cite{Song2016Quan}. In the CI phase, the interference pattern is highly asymmetric with respect to the field direction while the sinusoidal pattern persists, which reflects the broken time reversal symmetry in the presence of equilibrium magnetization (the MBT is a ferromagnet in the CI phase). This pattern is consistent with a recent experiment of the quantum Hall insulator~\cite{Amet2016Super}. These striking features in the interference patterns associated with the AI and CI phases are benchmarked against the ordinary metal phase [in Fig.~\ref{inter}(c)], which exhibits a typical Fraunhofer pattern where the width of the central peak around $\phi=0$ is doubled compared with the period in Figs.~\ref{inter}(a) and (b).

In summary, we have investigated the transport properties of a superconductor-MBT-superconductor Josephson junction and find that the system exhibits distinct quantum interference patterns in the presence of a perpendicular magnetic field, which are unique to different magnetic phases of the MBT. These interference patterns also reveal the spatial distributions of supercurrents in different phases and can be used to experimentally distinguish the versatile phases in the MBT thin film. 

Y.-H.L. acknowledges fruitful discussions with Hua Jiang. This work is supported by the Air Force Office of Scientific Research under Grant No. FA9550-19-1-0307.

\bibliography{ref}

\end{document}